\documentclass[12pt,a4paper]{article}
\pdfoutput=1
\newcommand{\be}{\begin{equation}}
\newcommand{\ee}{\end{equation}}
\newcommand{\bea}{\begin{eqnarray}}
\newcommand{\eea}{\end{eqnarray}}
\newcommand{\beas}{\begin{eqnarray*}}
\newcommand{\eeas}{\end{eqnarray*}}
\newcommand{\ba}{\begin{array}}
\newcommand{\ea}{\end{array}}
\newcommand{\nn}{\nonumber}

\newcommand{\bt}{\begin{table}}
\newcommand{\ve}{\varepsilon}
\newcommand{\vsi}{\varsigma}

\newcommand{\al}{\alpha}
\newcommand{\ga}{\gamma}

\newcommand{\Ga}{\Gamma}	
\newcommand{\dal}{{\partial^2}}
\newcommand{\de}{\delta}
\newcommand{\De}{\Delta}
\newcommand{\ka}{\kappa}
\newcommand{\la}{\lambda}
\newcommand{\La}{\Lambda}
\newcommand{\na}{\nabla}

\newcommand{\si}{\sigma} 
\newcommand{\Si}{\Sigma}

\begin{document}
\title{\bf 
Quartet-metric  general relativity:
scalar graviton,  dark  matter and  dark energy
 }
\author{Yury~F.~Pirogov
%\footnote{E-mail: yury.pirogov@ihep.ru }
\\
\small{\em 
%Theory Division,  
SRC  Institute for High Energy Physics of
NRC Kurchatov Institute, Protvino 142281, 
%Moscow Region,
 Russia }
%\\ \small{\em E-mail:  yury.pirogov@ihep.ru}
}
\date{}
\maketitle

\begin{abstract}
\noindent
General Relativity  extended through  a dynamical  scalar quartet
is proposed  as  a  theory of the scalar-vector-tensor gravity, 
generically describing the unified gravitational dark matter (DM) and dark
energy (DE).  The implementation in the weak-field limit  of the Higgs mechanism
for the gravity, with a redefinition of  metric field, is exposed in a
generally
covariant form.  Under a natural  restriction  on parameters,  the   redefined
theory  possesses in the linearized approximation  by  a  residual
transverse-diffeomorphism invariance, and consistently comprises   the massless
tensor graviton and  a  massive scalar one  as a DM particle.  
A number of  the adjustable parameters in the   full nonlinear
theory and a partial decoupling of the latter  from  its weak-field limit 
noticeably extend the perspectives for the unified description of the gravity DM
and DE in the various phenomena  at  the different scales.

\end{abstract}

\section{Introduction}

The unification of the superficially unrelated phenomena  in nature seems to be
the main trend in the contemporary fundamental physics. Among such the formally
unrelated  phenomena there are  the so-called dark matter (DM) and dark
energy (DE).  However,  both  DM and DE    being extremely 
elusive, they may naturally  have their common   origin in a 
modification/extension of  General Relativity (GR).
The latter   is well-known to be the generally covariant (GC) metric theory of
gravity describing  in vacuum one physical gravity  mode,  the massless
two-component transverse-tensor  graviton. It is commonly adopted  that GR is 
not an ultimate theory of gravity. Nevertheless, it well  may serve as a   firm
basis  for a  future (more) fundamental  theory.  In particular, among the
conceivable extensions to   GR one may distinguish that with an additional
(massive) scalar graviton. Such a GR extension, in comparison  with  a number
of other ones, will mainly be addressed to in what follows.\footnote{For a
review on the extended theories of gravity, see, e.g.,~\cite{Capo}.}

The  masslessness  of the tensor graviton   is safely  ensured  in   GR  by the
conventionally adopted  general gauge invariance/relativity.  At
that, a scalar gravity mode,  contained {\em a~priori}  in metric,   gets 
 unphysical as a by-product of the general gauge invariance/relativity.
However, to insure the generic property  of the tensor masslessness  it  would
suffice for a gravity theory to  possess the invariance just  under the
 volume-preserving/transverse diffeomorphisms (TDiff's)~\cite{Ng}.
Given  this,   there could appear in  metric  one more physical  gravity
mode,  the scalar one.  
A  theory of gravity  based  on  the explicit GR violation, with 
the residual TDiff  invariance and an 
extra scalar mode contained in metric,   was proposed   in the 
non-GC form in~\cite{Buch}--\cite{Creutz1} and further elaborated  
in~\cite{Alv}. In the explicitly GC    form, such a theory 
was proposed  in~\cite{Pir0}  and developed 
in~\cite{Pir0a}--\cite{Pir1}.  At that, the explicit 
GR  violation  was  treated as a {\em  raison d'\^etre} for  appearance of 
the gravitational DM.\footnote{The  term ``GC
violation'', used previously  in ~\cite{Pir0} and later,  is  more
appropriately substituted here   by the ``GR
violation''~\cite{Pir1}.} To such an  interpretation, GC preservation 
proves to be crucial.  The  emergent   scalar-graviton  DM proved to  possess 
the generic properties conventionally 
assigned to DM.  In particular, due  to the coherent field of scalar gravitons
there exists  in  vacuum a halo-type solution with a constant asymptotic
rotation velocity~\cite{Pir1a,Pir1}. 
Similarly to the black holes  (BH's)  serving as   a signature for the plain GR,
 the halo-type objects may serve as a signature for GR  extended through  the
scalar graviton. Associated with the {\em local scale} invariance, the
respective scalar graviton was called  previously a {\em 
systolon}~\cite{Pir1}. 
To ensure  the GC  preservation one should
inevitably introduce a nondynamical scalar density (of an unknown 
nature).\footnote{A nondynamical 
scalar  density for  a four-volume  element  appeared originally  in the 
Unimodular Relativity (UR)~\cite{Anderson}. Though inevitable in  UR for GC, 
such a scalar density  is nevertheless often missed, tacitly implying the
special coordinates, as well as  the absence  of the scalar-density 
singularity (not necessarily fulfilled in a more general case).} 
This is the simplest theory realizing the scenario of the explicit  GR
violation with the  gravity~DM. 

A  more detailed study  of  the gravity DM  may though 
require an extension   of the scenario beyond the
minimal one. Irrespective of DM, the general second-order
effective Lagrangian with the explicit GR
violation   in the non-GC form  was proposed in~\cite{Don}.
In the GC form,  such a Lagrangian 
was obtained  in the  context of a  nonlinear model in ~\cite{Pir2}.
Similarly to the minimal   case, to maintain GC under  the explicit  GR
violation  it is necessary to introduce a  nondynamical affine connection 
expressed  minimally   through a  quartet of the scalar fields. Such
a proliferation of the uncontrollable nondynamical quantities in a
semi-dynamical model makes one uneasy,  both phenomenologically
and theoretically,  and  may result   in some   conceptual
problems. It would thus  be desirable to make the  effective
field  theory of gravity   fully dynamical by converting the nondynamical
quantities, minimally the scalar quartet, to the dynamical ones. 
A dynamical  scalar quartet  in the context of  the  space-time four-volume
element was   proposed  originally in \cite{Guendel}. 
 For an  implementation  of the Higgs mechanism for the massive tensor gravity
it was   introduced in~\cite{Hooft}.
A special representation for the  gravity Higgs fields in  terms of  the 
scalar quartet  was proposed in  \cite{Cham} and  worked out in \cite{Oda}.
It  helps to solve some problems inherent  to  the massive modification of GR. 
Such a representation, with  a modification hereof,   is  used   for the   
consistent treatment of   a more critical  extension  to GR in the present
paper.
    
The  paper  develops  the preceding
results due  to the author on   the scalar-graviton/says\-tolon 
DM~\cite{Pir0}--\cite{Pir1}, modifying  the semi-dynamical model  to a fully
dynamical theory.  First of all, a general, quadratic
in the  derivatives of   metric,  gravity kinetic Lagrangian  in an explicitly 
GC form~\cite{Pir2}  is used instead of the minimal  one.  Besides,  the
explicit GR violation, with a nondynamical affine connection,  is  abandoned 
due to    a dynamical scalar  quartet. 
In  Section~2,   the full  nonlinear theory  of the quartet-metric (QM)
GR, or, for short,  the {\em quartet-metric gravity} (QMG) is considered.
The GC  frameworks  for  the unification of the  gravity  DM  and DE  are
worked out.  A coupled system  of the classical  field equations (FE's)  for 
metric and the scalar quartet is  presented. 
The minimal   GR extension, in comfort with  the previous
semi-dynamical  results~\cite{Pir1a,Pir1}, is considered in more detail. 
In Section~3, the weak-field  (WF) theory corresponding to the full nonlinear
one is considered.  The   implementation  of the Higgs
mechanism for the scalar-vector-tensor gravity  in the arbitrary metric and
quartet backgrounds is explicitly demonstrated in a GC form.
The  linearized approximation (LA) for the  most general version  of the theory,
as well as  for  its  natural   reduction  insuring TDiff invariance,   are 
then considered. The latter case,  being   unitary and   free of
ghosts and the classical instabilities,  is argued  to consistently comprise  
the massless tensor graviton and a massive scalar one as a DM particle.  The
nearest and far-away prospects for  QMG  are   shortly discussed  in Conclusion.

\section{Full nonlinear theory}

\subsection{Quartet-metric gravity}

An underlying theory of gravity and  space-time,   whichever it might
be, should inevitably manifest itself on an
observable level as an  effective field theory to match with GR and the
conventional field theory for the ordinary matter.  Thus in searching for  such
a  fundamental theory, one should, conceivably, look
first for the respective effective theory. The latter is to be generically
characterized  by a set of fields and symmetries ruling the interactions of the
fields. Let us thus assume   that the effective field theory of  the
extended gravity superseding GR    is   described by the  dynamical fields of 
metric $g_{\mu\nu}$ and a  scalar quartet ${Q}^a$, $a=0,\dots,3$, with an action
\be\label{MXL} 
S=\int L_G (g_{\mu\nu},   {Q}^a) \sqrt{-g} \, d^4 x,
\ee
where  $g=\det(g_{\mu\nu})$ and $L_G$ is an effective 
Lagrangian.\footnote{Accordingly, in the $d$ space-time 
dimensions there should take place  the {\em d-plet-metric  gravity}.}  
The  latter is assumed to have a GC form and to be  invariant under a  global
internal  Lorentz symmetry $SO(1,3)$ acting on the
indices $a,b$, etc, with  the invariant Minkowski symbol 
$\eta_{ab}$. It is  understood that the signature of $\eta_{ab}$ determines
the signature of $g_{\mu\nu}$. The physical meaning of the extra variables
${Q}^a$ will be clarified later on.  Here  we only mention  that  precisely  
these  variables are responsible for the  unified gravity DM
and DE. Without loss of generality, $L_G$ may
generically be decomposed into  three parts  depending on the appearance of the
derivatives of metric: 
\be
\label{LgX} 
L_G= L_g(\partial_\la g_{\mu\nu}  )  
+ \De  K ( \partial_\la g_{\mu\nu} ,   {Q}^a)-  V ( {Q}^a).
\ee
An additional dependence directly
on $g_{\mu\nu}$ to ensure GC is tacitly allowed, too.   
The part $L_g$ is a   Lagrangian of the pure-metric gravity, with $\De K$ and 
$V$ meaning,  respectively,  the {\em hard}/kinetic   and  {\em
soft}/potential  admixtures to  the pure-metric gravity.
The order of the derivatives of ${Q}^a$ depends on the context 
(see, later).  At that, the appearance of ${Q}^a$  without  derivatives
is forbidden due to  the assumed  shift  symmetry ${Q}^a \to {Q}^a +
C^a$, with the  arbitrary constants $C^a$. 
The more tightly localized kinetic contributions are to be
 associated with the gravity DM, while the more loosely distributed potential
ones with the gravity DE, roughly in compliance with their observational
abilities to  the  spatial clusterization. 
The gravity DE and DM may thus be  treated as
the two facets  of a single  {\em dark substance} (DS), with 
the phenomenon of the  unified origin  of  such  a DS 
due to the  common ${Q}^a$   referred to as the {\em dark unification}.
Besides, after  choosing a  classical background and expanding  the fields
around the background,  the scalar-quartet  admixtures result in the appearance
of the extra gravity degrees of freedom (d.o.f.'s) presenting the DS particles.
Consider  the various contributions to $L_G$  in more detail.

\subsection{Pure-metric gravity}

Similarly  to  the plain GR,    one  may    take  for the pure-metric  gravity
the minimal  GC  Lagrangian of the second order  in the derivatives
of metric:
\be 
L_g= - \frac{\ka_g^2}{2} R.
\ee
Here  $R$ is the Ricci scalar curvature, $\ka_g=1/(8\pi G_N)^{1/2}$ the
Planck mass and  $G_N$ the Newton's
constant. In what follows, we put $\ka_g=1$.  A   GR modification, with a GC
$L_g$  dependent only on metric, e.g., $f(R)$,  is {\em a  priori}
conceivable,~too. 
The more   crucial   GR extensions  are due to the  admixtures of
${Q}^a$ considered below.

\subsection{Hard/kinetic extension}

This kind of the GR  extension  is produced by
the effective operators of   the second order in  the metric derivatives,
likewise $L_g$. Instead of the derivatives of    metric, one may equivalently
use
the Christoffel connection built of $g_{\mu\nu}$:
\be\label{Ga}
\Ga^\la_{\mu\nu}=\frac{1}{2}g^{\la\ka}(\partial_\mu g_{\nu\ka}+ \partial_\nu
g_{\mu\ka} -
\partial_\ka g_{\mu\nu}),
\ee 
so that $\partial_\nu g_{\mu\rho}=  g_{\rho\la} \Ga^\la_{\mu\nu}  +   g_{\mu\la}
\Ga^\la_{\rho\nu}$.
To preserve GC, the kinetic  effective Lagrangian  $\De K$  should
depend on the difference of   $\Ga^\la_{\mu\nu}$  and an auxiliary  
 affine connection~$  {\ga}^\la_{\mu\nu}$ (for simplicity,
symmetric)~\cite{Pir2}:
\be\label{B} 
B_{\mu\nu}^{\la}=\Ga_{\mu\nu}^\la - {\ga}_{\mu\nu}^\la.
\ee
Such an  auxiliary connection may minimally  be  taken as follows:
\be\label{{Q}} 
{\ga}_{\mu\nu}^\la =\frac{ \partial^2 
{Q}^{a}}{\partial x^\mu\partial x^\nu} 
 \frac{ \partial x^\la}{\partial {Q}^{a}}\bigg|_{{Q}^a= {Q}^a(x)}
 \equiv {Q}^\la_{a} \partial_\mu{Q}^a _\nu,
\ee
where  ${Q}^a_\mu\equiv \partial_\mu {Q}^a$ and  ${Q}^\mu_a\equiv  \partial
x^\mu/\partial  {Q}^{a}|_{{Q}^a= {Q}^a(x)}$, with ${Q}\equiv \det (\partial
{Q}^a/\partial x^\mu ) \neq 0$  or $\infty$ for the invertibility,
$ x^\mu = x^\mu({ {Q}^a})$.  It is the latter
requirement which picks out $Q^a$ from other conceivable scalar fields. 
Assign  to   the extra variables ${Q}^a$   the dimension
of length and  the  meaning as follows.   Namely, assume  that the
vacuum is a kind of a physical medium  modelled  (ideally) by an
affine-connected  (i.e., possessing by an affine connection) 
manifold endowed  with the  {\em absolute/affine} (i.e., defined modulo the
Lorentz transformations and translations) coordinates  ${Q}^a$.
At that, the conventional GR deals  only  with the {\em relative/observer's}
coordinates $x^\mu$ undergoing the arbitrary (continuous)  transformations among
themselves.\footnote{In a
sense, it  is proposed  here  a merging, at a next level,  of the Newtonian and
GR approaches to gravity and space-time.}  The quantities ${q}^\al\equiv
\de^\al_a
{Q}^a $, insuring  
${\ga}^\ga_{\al\beta}({q})\equiv0$,   define  the 
distinguished  observer's coordinates  coinciding with  $Q^a$.
With respect to the arbitrary $x^\mu$, the distinguished ${q}^\al$  are the
dynamical quantities, with   $  {Q}^{\al}_\mu=\partial   {q}^\al /\partial
x^\mu$  and ${Q}^\mu_{\al}= \partial x^\mu /\partial {q}^\al|_{q^\al=q^\al(x)}$
being  the frames relating these  coordinates.  In reality, 
the transition between the two kinds  of coordinates  may be  singular on a set
of points (including, generally, the infinity). The extension of the set of
the  dynamical variables beyond the metric is tamed eventually by the
expansion of the general gauge invariance/relativity on the whole this set,
leaving  the  total number of the independent field  variables still equal to
ten (as
in  the metric). 

In the presence of  $Q^a$, one  may substitute
a non-metrical affine connection    ${\ga}^\la_{\mu\nu}$ by a
Christoffel one, ${\ga}^\la_{\mu\nu}
=\Ga^\la_{\mu\nu}({\ga}_{\ka\rho})$, similarly to (\ref{Ga}).
It corresponds to  an auxiliary metric (the {\em affine metric}):
\be 
{\ga}_{\mu\nu}={Q}^a_\mu {Q}^b_\nu\eta_{ab},
\ee
with  the inverse  ${\ga}^{\mu\nu}={Q}_a^\mu
{Q}_b^\nu\eta^{ab}$, and ${\ga}\equiv\det({\ga}_{\mu\nu})=-{Q}^2\neq 0$ or
$\infty$  for the invertibility. 
By this token,  formally ${Q}_\mu^a= {\ga}_{\mu\la} \eta^{ab }{Q}^\la_b$ and
${\ga}_{\mu\nu}= {\ga}_{\mu\ka}{\ga}_{\mu\la}{\ga}^{\ka\la}$. In the
distinguished coordinates~${q}^\al$,  the affine metric coincides with the
Minkowski symbol, $\ga_{\al\beta}=\eta_{\al\beta}$ (simultaneously  with
${\ga}^\ga_{\al\beta}=0$).\footnote{Stress that
QMG is  basically a one-metric theory. Using   the  affine metric
${\ga}_{\mu\nu}$, instead of ${Q}^a$,   is not obligatory,  though 
 technically convenient and geometrically clarifying.}

Being given by the  difference between  the  similarly   transforming
quantities, the field $B_{\mu\nu}^{\la}$ is a  true  tensor
and, as such, may serve to construct a GC scalar Lagrangian.
Introducing a complete  set of  
the  independent    partial  kinetic operators, bilinear in $ B_{\mu\nu}^\la$:  
\bea\label{O_p}
 {K}_1= g ^{\mu\nu} B_{\mu\ka}^\ka B_{\nu\la}^\la,   && 
 {K}_2=g_{\mu\nu}  g^{\ka\la}g^{\rho\si}
B_{\ka\la}^\mu   B_{\rho\si}^\nu,\nn\\
 {K}_{3}=g^{\mu\nu} B_{\mu\nu}^\ka B_{\ka\la}^\la,      &&  
{K}_4 = g_{\mu\nu}  g^{\ka\la}  g^{\rho\si} B_{\ka\rho}^\mu
B_{\la\si}^\nu    ,\nn\\  
 {K}_5=  g^{\mu\nu}  B_{\mu\ka}^\la  B_{\nu\la}^\ka,   && 
\eea
decompose $\De K$  as 
\be\label{VUR} 
\De K = \frac{1}{2} \sum_{p=1}^5   \ve _p   {K}_p ,
\ee
with some  free parameters $\ve_p$,  $p=1,\dots, 5$,  presumably  small,
$|\ve_p|\ll 1$.  The two  more second-derivative  linear in $B^\la_{\mu\nu}$
terms,  $g^{\mu\nu} \na_\la B^\la_{\mu\nu}$ and $g^{\mu\nu} \na_\mu
B^\la_{\nu\la}$, with $\na_\la$
a covariant derivative, are  omitted  due to the imposed  invariance  under
the reflection $ B_{\mu\nu}^\la\to -  B_{\mu\nu}^\la$.  This is
the   fully dynamical GR development of a semi-dynamical  approach,  the latter
based on  the explicit  GR violation and a nondynamical
$\hat{\ga}^\la_{\mu\nu}$~\cite{Pir2}.

\subsection{Soft/potential  extension}

Such a GR  extension, implementing the Higgs mechanism for gravity,  is produced
by the effective operators, potentials,   containing metric only without 
derivatives. To this
end, take as the scalar  fields the (dimensionless)
$SO(1,3)$ decouplet~\cite{Cham} 
\be\label{Si} 
\Si^{ab}=g^{\mu\nu} {Q}^a_\mu  {Q}^b_\nu
\ee
incorporating   the  singlet 
\be
\Si= \eta_{ab}\Si^{ab} = {\ga}_{\mu\nu}
g^{\mu\nu}. 
\ee
Substitute  them  equivalently  by  
\bea
\tilde\Si^{ab}&\equiv&\Si^{ab}- \frac{1}{4}\eta^{ab}\Si ,\nn\\
\Si_0&\equiv& \Si -4,
\eea
so that  $\tilde \Si\equiv\eta_{ab}\tilde\Si^{ab}=0$. 
Supplement $\Si_0$   by one more
(dimensionless)  scalar field 
\be\label{si} 
\si =-\frac{1}{2} \ln |\det(\Si^{ab})| =\ln \sqrt{-g}/|{Q}|   =\ln \sqrt{-g}/
\sqrt{-{\ga}}   .
\ee
The latter  field  is  precisely the one  used previously for the
scalar graviton/systolon~\cite{Pir0}--\cite{Pir1}.
The  fields $\tilde \Si^{ab}$,  $ \Si_0$  and $\si$ prove to be 
homogeneous  linear in the weak metric field  in LA (see, later) and   are thus
suitable for building  a perturbative potential without a cosmological
constant. The latter  may be added, if desired, explicitly. 
Without loss of generality, the potential may  be presented as follows:
\bea\label{V} 
V &=&     \frac{1}{8}m_t^2 \Big( \tilde\Si^{ab} \tilde
\Si_{ab} - \frac{3}{4}  \Si_0^2\Big ) +  \frac{1}{8}m_0^2\Si_0^2\nn\\
&& \pm \  \frac{1}{4}m_x^2 \Si_0\si+ \frac{1}{2} m_\si^2 \si^2
+\De V( \tilde \Si^{ab},    \Si_0, \si ),
\eea
with    $m_t$,  $m_0$, $m_x$  and  $m_\si$  some mass
parameters,   and $\De
V$   a  rest of the potential comprising the  higher degrees of 
$\tilde\Si^{ab}$, $  \Si_0$  and~$\si$.
The particular  form of  the  first   term of $V$   and the meaning of
the parameters are to be justified by  the   compliance  with  the Fiertz-Pauli
LA Lagrangian   (see, later).  The putative terms dependent only on 
${\ga}_{\mu\nu}$, such as $R({\ga}_{\mu\nu})$, are absent due to the
affine  flatness of the space-time  manifold, with ${\ga}_{\mu\nu}$ reducing
in the  distinguished  coordinates~${q}^\al$ to
the Minkowskian $\eta_{\al\beta}$,  hence $R(\ga_{\mu\nu})=0$.

\subsection{General diffeomorphism  invariance}

 Ultimately, the independent variables  of the theory are  $g_{\mu\nu}$ and
${Q}^a$. The Lie derivatives,  defining  the  {\em dynamical/active} field 
 transformations corresponding  the change of coordinates  
${\cal D}_\xi x^\la =- \xi^\la$, with $\xi^\la$ an arbitrary shift  vector, are
as follows:
\bea\label{GenDiff} 
{\cal D}_\xi {Q}^a&=&\xi^\la \partial_\la {Q}^a \equiv\xi^\la  {Q}^a_\la ,\nn\\ 
 {\cal D}_\xi {Q}^a_\mu = \partial_\mu {\cal D}_\xi {Q}^a&=& {Q}^a_\la
\partial_\mu
\xi^\la  + \xi^\la \partial_\la   {Q}^a_\mu,\nn\\
 {\cal D}_\xi {Q}_a^\mu&=&- {Q}_a^\la
\partial_\la \xi^\mu  + \xi^\la \partial_\la   {Q}_a^\mu,\nn\\
{\cal D}_\xi g_{\mu\nu} &=& g_{\mu\la} \partial_\nu \xi^\la  + g_{\nu\la}
\partial_\mu \xi^\la  +\xi^\la \partial_\la g_{\mu\nu}, \nn\\
{\cal D}_\xi g^{\mu\nu} &=& -g^{\mu\la} \partial_\la \xi^\nu - g^{\nu\la}
\partial_\la \xi^\mu  +\xi^\la \partial_\la g^{\mu\nu}
\eea
(and similarly for ${\ga}_{\mu\nu}$ and   ${\ga}^{\mu\nu}$). Henceforth, one
can get the Lie derivatives of other quantities,  e.g., 
\be
{\cal D}_\xi \sqrt{-g}=\partial_\la(\sqrt{-g}\xi^\la)
\ee
(and similarly for $\sqrt{-{\ga}}$), 
with  $\si$ transforming thus as a scalar (as it should):
\be
{\cal D}_\xi \si = \xi^\la \partial_\la \si. 
\ee
The same concerns the scalars $\Si^{ab}$ and $\Si$. 
The  Lie derivative of any quantity  may explicitly be  expressed in a tensor
form through  replacing $\partial_\mu$ by a 
covariant derivative  $\na_\mu $.  In particular, one gets ${\cal D}_\xi
g_{\mu\nu} =\na_\mu \xi_\nu +\na_\nu \xi_\mu$, where $\na_\la g_{\mu\nu}=0$ and
$\xi_\mu=g_{\mu\la} \xi^\la$, etc.
Due to GC,   the full nonlinear theory, as  containing only the dynamical
fields,  is automatically  gauge invariant under the general  diffeomorphisms
(GDiff's).\footnote{This is what distinguishes   the fully
dynamical theory from a semi-dynamical model, where a residual  gauge
invariance/relativity is determined, under GC,  through fixing the
nondynamical fields~\cite{Pir1}.}
The latter ones  reduce the number of the independent field components in
$L_G$  up  to  ten (vs.\ six in GR). To account for the gauge degeneracy, 
a  gauge fixing Lagrangian,  $L_F$,
appropriate for the  problem at hand, is to be added. 
In particular, one may impose  the same gauge conditions on the metric alone as
in GR.

\subsection{Extended classical field equations}

Supplementing the  gravity Lagrangian $L_G $ by  the  ordinary  matter one, 
$L_m$,  and varying the total action with respect to $\de g^{\ka\la}$
and $\de {Q}^a$   ($\de {Q}^a_\mu =\na_\mu \de {Q}^a$), so that, in
particular: 
\bea 
\de \Si^{ab}&=&   {Q}^a_\ka {Q}^b_\la \de g^{\ka\la}+ 2 g^{\ka\la}{Q}^a_\ka \de
{Q}^b_\la,\nn\\
\de  \Si&=&\eta_{ab} \de \Si^{ab}   = {\ga}_{\ka\la}\de g^{\ka\la}+  g^{\ka\la}
\de{\ga}_{\ka\la} ,\nn\\
\de   {\ga}_{\ka\la}  &=&  \eta_{ab}({Q}^a_\ka \de  {Q}^b_\la  +{Q}^a_\la  \de
{Q}^b_\ka), \nn\\
\de\sqrt{- {\ga}}&=& (1/2)\sqrt{-{\ga}}{\ga}^{\ka\la}\de {\ga}_{\ka\la}\nn\\
\de\sqrt{- g}&=&- (1/2)\sqrt{-g} g_{\ka\la}\de g^{\ka\la},\nn\\
\de \si&=&  \de\sqrt{- g}/\sqrt{- g}  -   \de\sqrt{- {\ga}}/\sqrt{- {\ga}},
\eea
with $\de g_{\mu\nu}=-
g_ {\mu\ka} g_{\nu\la} \de g^{\ka\la}$ (and similarly for $\ga_{\mu\nu}$),  one 
gets a pair of the coupled classical FE's  for QMG
 in a generic form as follows: 
\bea\label{FE} 
G_{\mu\nu}\equiv 
R_{\mu\nu}-\frac{1}{2} R g_{\mu\nu}  & = & 
 T_{\mu\nu} ^K  + T_{\mu\nu}^V   +  T_{ \mu\nu}^m   \equiv
 T^D_{\mu\nu} + T^m_{\mu\nu} ,\nn\\
 \na_\ka   (\De  K^{\ka\la} {Q}^a_\la )&=& 
\na_\ka\bigg(\Big(\frac{\partial V}{\partial \Si^{cb}}  g^{\ka\la}\eta^{ca}
-(\frac{1}{2}\frac{\partial  V}{\partial \si}  {\ga}^{\ka\la} +
\frac{\partial L_m}{\partial{\ga}_{\ka\la}} ) \de^a_b \Big)       
{Q}^b_\la\bigg),
\eea   
where $\De   K^{\ka\la}\equiv  \de\De  K/\de  {\ga}_{\ka\la}$, with $ \de/\de
{\ga}_{\ka\la}$  a total variational derivative with respect
to $  {\ga}_{\ka\la} $.  
The l.h.s.'s of FE's depend on the second derivatives of metric, while the
r.h.s.'s only on the first derivatives. 
In the spirit of DM, $L_m$  is assumed to  depend on ${Q}^a$
(if any) exclusively through ${\ga}_{\ka\la}$, without its derivatives.
To eliminate the gauge ambiguity, in  solving FE's one should 
first fix the coordinates by imposing   an appropriate  gauge
condition, which  will tacitly be understood. In reality, fixing the coordinate
to   maximally  simplify  the metric  FE's   proves to be the  most
appropriate.\footnote{On the contrary, choosing the distinguished  coordinates
$q^\al$, superficially convenient,  one
would   loose such a  freedom of simplifying the metric FE's.}
The l.h.s.\ in the upper line  of (\ref{FE}) is the gravity tensor 
$G_{\mu\nu}$ due to $L_g$, with $T^D_{\mu\nu}\equiv
\De T^K_{\mu\nu}+T^V_{\mu\nu}$ in  the r.h.s.\  treated  as the
energy-momentum tensor of DS. This is, in essence, the {\em raison
d'\^etre} for associating the admixtures due to ${Q}^a$ with DS. 
The kinetic contribution to $T^D_{\mu\nu}$~is
\be 
\De T_{\mu\nu}^K\equiv
\frac{1}{2} \sum_{p=1}^5
\ve_p T^K_{p\, \mu\nu}
\ee
where $T^{K}_{p\,\mu\nu}$ are  the partial contributions due to  ${K}_p$: 
\be 
T^K_{p\, \mu\nu}=\frac{2 } {\sqrt{-g}    }\frac{\de\Big (\sqrt{-g}
{K}_p\Big)}{\de g^{\mu\nu}}.
\ee
with $\de/\de g^{\mu\nu}$ designating  a
total variational derivative with respect to $g^{\mu\nu}$.
A similar expression holds for    the canonical
energy-momentum tensor   $ T^m_{ \mu\nu}$ of
the ordinary matter.
Likewise, the potential   contribution  to  $T^D_{\mu\nu}$  is
\be 
T_{\mu\nu}^V=-2 \frac{\partial V}{\partial \Si^{ab}}{Q}^a_\mu {Q}^b_\nu 
+  \Big (\frac{\partial V}{\partial \si}  +V \Big) g_{\mu\nu}.
\ee
In particular, a nonzero constant part of $V$ would  correspond to a
cosmological term.
Due to the reduced  Bianchi identity,  $\na_\mu  G^{\, \mu\nu}=0$,    the total
energy-momentum tensor, $T_{\mu\nu}$,   
should be  covariantly conserved:
\be 
\na_\mu  T^{\mu\nu}\equiv  \na_\mu ( T_m^{\mu\nu}  + T_D^{\mu\nu} )= 0.
\ee
Assuming $L_m$ to be  independent of ${Q}^a$,  one conventionally
gets with account for the ordinary matter FE's  that  $\na_\mu T_m^{\mu\nu}=0$.
 In this case (or in the  matter vacuum),  the DS contribution  should
separately  be  covariantly conserved. 

The  system of FE's (\ref{FE}) presents  an  essence of the fully dynamical 
theory. Namely, it may be said that~(\ref{FE})  determines  in a
self-consistent  manner a  two-level structure of the space-time manifold:  a
basic  affine  structure  and a fine metric one. The first line
of FE's~({\ref{FE}), obtained by varying only $g_{\mu\nu}$,  remains the same
independently of whether ${Q}^a$ are dynamical or not. 
Thus  the first  equation~({\ref{FE}),   determining a  metric structure  at a
given  affine one, is  unchanged compared to  the
semi-dynamical model.  The second equation  acts  v.v., determining the
back reaction of the  metric structure on the affine one.  At the very
least, a self-consistent  solution  may  be looked-for  by means of  the
consecutive approximations starting from a putative  solution (under convergence
of the procedure). The account for the full dynamics should  
restrict a prior freedom of choosing  an otherwise arbitrary  nondynamical 
background  in the  semi-dynamical approach.
For a special case, this is worked-out  below.

\subsection{Minimal  extension}

As a   minimal kinetic  contribution  to the quartet GR  extension, one 
may  consider  the operator
\be 
{K}_1= g^{\mu\nu}\partial_\mu\ln \frac{\sqrt{-g}}{  |{Q}|  }
\partial_\nu\ln\frac{\sqrt{-g}}{  |{Q}|   }. 
\ee
where  $|{Q}|\equiv|\det({Q}^a_\mu)|=\sqrt{-{\ga}}$.
This restricted case  presents a fully dynamical generalization of  the
semi-dynamical  model  for  the scalar-graviton/systolon  DM, with  a
nondynamical $\hat {\ga}$
and  the explicit GR  violation~\cite{Pir0}--\cite{Pir1}.  
Under  the GR extension exclusively  through ${K}_1(\si)$ and $V(\si)$, there
is a parity between the numbers of the 
 independent variables in the full nonlinear theory and  in its LA,
both being seven (see, later).
The quartet enters only through $\sqrt{-{\ga}}=|{Q}|$,   irrespective  of
$g_{\mu\nu}$. Under the metric variation, ${\ga}$ remains  thus unvaried. Hence
the  scalar-graviton  part of the metric FE's in vacuum,  obtained  previously 
in the semi-dynamical model~\cite{Pir1a,Pir1}:
\be\label{FEsi}
\ve_1\na^\ka\na_\ka \si +\partial
V/\partial \si = \La_0 e^{-\si} , 
\ee
remains  unchanged. In the above,  $\La_0$ is an integration constant arising
due to  the reduced  Bianchi identity.
On the other hand, with $K_1$  dependent  on
${Q}^a$ only
through  $\si$,   the quartet  part of FE's in vacuum becomes  as follows: 
\be 
\na_\la\Big((\ve_1 \na^\ka\na_\ka \si
+\partial V/\partial \si) {Q}^\la_a\Big) =0.
\ee
Combining the two equations one gets  the consistency condition:
\be\label{consist} 
\La_0 \na_\la (e^{-\si}{Q}^\la_a) = ( \La_0/\sqrt{-g}) \partial_\la  (\sqrt{-g} 
e^{-\si}{Q}^\la_a)= 0.
\ee
Several  static spherically symmetric solutions of the metric part of FE's 
at $V=0$ in the semi-dynamical  model are presented in~\cite{Pir1}.
Their prolongation  to the fully dynamical  theory is as  follows.

\paragraph{Extended cosmic objects}

Let first  $\La_0\neq 0$.   This case corresponds  to  the extended cosmic 
objects in the vacuum,   the {\em  dark halos} (DH's),   peculiar exclusively to
QMG (or to its semi-dynamical counterpart).
 Under a regular boundary condition in the spatial origin,   ${\ga}$ is
determined in the
chosen  (due to  fixing a gauge for metric) observer's coordinates $x^\mu$ from
the first part of  FE's  (\ref{FE}) by the metric  and $\si$,  as in the
semi-dynamical model~\cite{Pir1a,Pir1}. 
In view of (\ref{si}),  $e^{-\si}=\sqrt{-{\ga}}/\sqrt{-g}$, 
Eq.~(\ref{consist}) gives 
\be\label{quart}
\partial_\la (\sqrt{-{\ga}} {Q}^\la_a)=0,
\ee
with the metric $g_{\mu\nu}$  falling-off. 
We consider  the  static spherically  symmetric  solutions to the  metric
FE's, with $\ga=\ga(r)$, $r$ the distance from the origin.
Transforming the expressions from the original coordinates  $x^\mu=(x^0,x^m)$, 
$m=1,2,3$, to 
the distinguished ones  $q^\al =(q^0, q^m)$, given by
$q^0=\sqrt{-\ga} x^0$,  $q^m= x^m$,
wherein $\ga=-1$,  one gets the solution  $Q^a=\de^a_\al q^\al$,
implying in  the previously fixed  coordinates $x^\mu$ the scalar quartet
$Q^a=(Q^{(0)} , Q^{A})$:
\be
Q^{(0)}=\sqrt{-\ga} x^0, \ \ Q^{A}  =  \de^{A}_k x^k,
\ee
where $A=1,2,3$.
For $\partial_\mu Q^a\equiv Q^a_\mu= (Q^{(0)}_\mu , Q^{A}_\mu) $ one thus gets
\bea
Q^{(0)}_0=\sqrt{-\ga},&&     Q^{(0)}_m=\frac{-\ga'}{2\sqrt{-\ga}}\,x^0 n_m,\nn\\
Q^A_0=0,&&Q^A_m=\de^A_m, 
\eea
and then for   its inverse  $Q_a^\mu= (Q_{(0)}^\mu , Q_{A}^\mu) $:
\bea
Q_{(0)}^0=\frac{1}{\sqrt{-\ga}},&&Q^m_{(0)}=0,\nn\\
Q^0_A=\frac{-\ga'}{2\ga}\,x^0\de_A^k n_k,&&Q_A^m=\de_A^m, 
\eea
where  $\ga' =d \ga/ d r$, $n^m=x^m/r$ and   $n_m=\de_{mk} n^k$. 
Eq.~(\ref{quart}) is satisfied, indeed.
Similarly, the affine metric  $\ga_{\mu\nu}= Q^a_\mu
Q^b_\nu\eta_{ab}$  in the original coordinates is as follows: 
\bea
\ga_{00}&=&-\ga,  \ \, \ga_{m0}= \ga_{ 0m}=  -\frac{1}{2} \ga' x^0n_m ,\nn\\
\ga_{ml}&=&-\de_{ml} -  \frac{1}{4 \ga} \ga'^2 (x^0)^2  n_m n_l
\eea
with  $\det(\ga_{\mu\nu}) = \ga$, indeed. 
 In the arbitrary $x^\mu$, $Q^a$ 
can then be found according to the transformation law for scalars.
Clearly, in the distinguished coordinates $q^\al$, where
$\ga=-1$,  the affine metric reduces to $\ga_{\al\beta}=\eta_{\al\beta}$, the
Minkowski symbol, with $\ga^\ga_{\al\beta}=0$. 
The DH space-time possesses thus by the flat affine structure  as  the
quasi-Euclidean  space-time, but,   in distinction with the latter, by  a
non-flat  metric structure.  The  principle difference  between the two
space-times is due to a singularity of DH's at  the spatial infinity.
The metric   structure,   obtained  in the semi-dynamical model,  remains still
valid (supplemented by the proper  affine structure) in the fully dynamical
theory. Hence, all the properties of   the galaxy  DH's, built exclusively of
the scalar gravitons/systolons as DM,  arrived at previously, remain  in
force.  With account for the asymptotic $1/r$-behaviour of the
attractive force in DH's~\cite{Pir1a,Pir1}, the latter ones  constitute   the
separate asymptotically confined ``mini-universes'' (in neglect by the edge
effects).

\paragraph{Compact cosmic objects}

Let  now $\La_0=0$.  In this case,    ${Q}^a$  remains still  unrestricted
except for a singular in   the  spatial origin $|{\rm det}
(Q^a_\mu)|=\sqrt{-\ga}$.
The latter is given through metric and $\si$ by an
exact solution to   the first   part of    FE's~(\ref{FE}) under the
regular  at the spatial infinity boundary conditions, similarly to the
semi-dynamical model.  The proper affine structure  can be  chosen to be
$\de$-wise flat (i.e., flat with exclusion of the  time-like  line  spreading
through the spatial origin).\footnote{Conceivably, this ambiguity may be
eliminated by treating DF's  as a limiting case of the more general cosmic
objects (see, next).} Hence, a wide class of the compact cosmic 
objects,   filled with  the scalar gravitons/systolons as DM, found  in the
semi-dynamical model remains still appropriate in the fully dynamical
theory.  In distinction with BH's of GR,  these objects are allowed
even in the absence of the ordinary matter, being caused by singularities of the
space-time itself  (henceforth   the  term a {\em dark
fracture} (DF)~\cite{Pir1}).  Modifying   BH's  of
GR,  DF's may possess by  a quite different structure of the event
horizon.    Note, that  DF's can still be
mimicked by BH's of GR with a (massless) scalar field (though,  of unknown
nature). One more kind of the cosmic objects in vacuum peculiar, as DH's, 
exclusively to QMG (and to its semi-dynamical counterpart) is as~follows.

\paragraph{Compact-extended cosmic objects}

At $\La_0\neq 0$,  there exist the  (approximate) vacuum  solutions to  the
extended FE's,  singular  in the spatial origin and at the spatial infinity, and
possessing, conceivably,
by  a
$\de$-wise flat affine structure. These  solutions    present  the very
peculiar  cosmic 
objects,  the  {\em dark lacunas} (DL's)~\cite{Pir1}, with the compact cores
and   extended tails, interpolating between DF's  and DH's. 
Having  DF's in their origin,   DL's  could model  the   galaxies (poor of the
ordinary matter).                                                           
Studying  such  the cosmic objects (and their matter, rotation and other 
modifications) to model  the real galaxies could present a future
challenge for~QMG.

\section{ Weak-field  theory}

\subsection{Weak-field limit}

\paragraph{Weak-field expansion}

The physics content of a nonlinear field   theory, as the quantum one, is
reflected  by its WF  limit. To this end,   consider the expansion
around some backgrounds  $ \hat {Q}^a$ and $ \bar g_{\mu\nu}$: 
\bea \label{small}
g_{\mu\nu}&=& \bar g_{\mu\nu} +h_{\mu\nu},\nn\\ 
{Q}^a&=&\hat {Q}^a+\chi^a,\nn\\ 
{Q}^a_\mu&=& \hat {Q}^a_\mu+\partial_\mu \chi^a,\nn\\
{Q}_a^\mu&=& \hat {Q}_a^\mu- \eta_{ab}\hat \ga^{\mu\la}\partial_\la \chi^b,
\eea
with    $ \hat {Q}^a_\mu=\partial_\mu \hat Q^a$,  $\hat Q^\mu_a$ the inverse to
$\hat Q_\mu^a$, $\hat \ga^{\mu\nu} = \hat Q^\mu_a \hat Q^\nu_b\eta^{ab}$,  
$g^{\mu\nu}= \bar g^{\mu\nu} -h^{\mu\nu}$, 
$h^{\mu\nu}=\bar g^{\mu\ka}  \bar g^{\nu\la}h_{\ka\la}$, 
and $|h_{\mu\nu}|, |\chi^a|\ll1$.
By default, the indeces  in the WF  limit 
are raised and lowered   by $\bar g^{\mu\nu}$ and
$\bar g_{\mu\nu}$, respectively, until stated otherwise. 
The parts $\hat {Q}^a$ and
$\chi^a$ may be associated  with a mean  value  and the  
fluctuations of the absolute/affine coordinates $Q^a$  (or  thus
the distinguished   $q^\al =\de^\al_a Q^a$) relative to  the smoothed 
observer's ones~$x^\mu$.  The latter coordinates  are  assumed to be fixed by a 
suitable gauge condition.  Ultimately,   $\hat {Q}^a$ and $\chi^a$ determine,
respectively, the classical and quantum manifestations of~DS.
In these terms, one~has
\be 
\Si^{ab}= \hat {Q}^a_\ka \hat {Q}^b_\la  (\bar g^{\ka\la} -  h^{\ka\la}) 
+\bar g^{\ka\la}( \hat {Q}^a_\ka \partial_\la \chi^b +  \hat {Q}^b_\ka
\partial_\la \chi^a).
\ee
To clarify the space-time structure of the WF theory in a GC form  it is more
appropriate to deal  exclusively with the space-time
notation. To this end,  introducing    $\chi^\mu \equiv \hat {Q}^\mu_a \chi^a$
($\chi^a =\hat {Q}_\ka^a \chi^\ka$) one gets for 
$\Si^{\mu\nu}\equiv \hat {Q}^\mu_a \hat {Q}^\nu_b \Si^{ab}$ ($\Si^{ab}=
\hat {Q}_\ka^a \hat {Q}_\la^b \Si^{\ka\la})$:
\be 
\Si^{\mu\nu}= \bar g^{\mu\nu} - h^{\mu\nu} +  \bar g^{\mu\la}
\hat\na_\la \chi^\nu +      \bar g^{\nu\la} \hat\na_\la \chi^\mu,    
\ee
where  
\be
\hat \na_\la \chi^\mu\equiv(\de^\mu_\ka\partial_\la  + \hat {Q}^\mu_a
\partial_\la \hat {Q}^a_\ka)\chi^\ka
\ee
is nothing but a covariant derivative with
respect to $\hat {\ga}_{\mu\nu}$, so that
$\hat \na_\la \hat {\ga}_{\mu\nu}=0$ (but not for
$\bar g_{\mu\nu}$).  
Similarly, for
\be
\ga_{\mu\nu}=\hat Q^a_\mu \hat Q^b_\nu\eta_{ab}+(\hat Q^a_\mu \partial_\nu
\chi^b+ \hat Q^a_\nu \partial_\mu \chi^b)\eta_{ab}
\ee
one gets 
\be
{\ga}_{\mu\nu}=\hat {\ga}_{\mu\nu} + \hat {\ga}_{\mu\la} \hat\na_\nu \chi^\la+
\hat{\ga}_{\nu\la} \hat\na_\mu \chi^\la,
\ee
where $\hat {\ga}_{\mu\nu} = \hat {Q}^a_\mu  \hat {Q}^b_\nu \eta_{ab}$, with  
$\hat  \ga^{\mu\nu}=\hat Q^\mu_a \hat Q^\nu_b \eta^{ab}$   being its  inverse.

\paragraph{Higgs mechanism for gravity}

Now, consider some arbitrary  backgrounds $\bar g_{\mu\nu}$
and $\hat Q^a$ (or, equivalently, $ \hat \ga_{\mu\nu}$) with the  coordinates 
fixed by a background gauge condition. Under the additional 
infinitesimal coordinate  transformations through  an arbitrary gauge
parameter $\zeta^\la$,  ${\cal D}_{\zeta} x^\la=
-\zeta^\la$, leaving by construction the backgrounds invariant,  
 one can find from
(\ref{GenDiff}) and (\ref{small})  for  $h_{\mu\nu}=\bar g_{\mu\ka} \bar
g_{\nu\la} h^{\ka\la}$ and $\chi^\mu$ the ensuing gauge transformations
\bea\label{gt} 
{\cal D}_{\zeta} h_{\mu\nu} & = &   \bar\na_\mu \zeta_\nu+  
\bar  \na_\nu \zeta_\mu ,\nn\\
{\cal D}_{\zeta}  \chi^\mu&=&\zeta^\mu.
\eea
Here  $\zeta_\mu=\bar g_{\mu\la} \zeta^\la$ and $\bar\na_\la$ is a covariant
derivative with respect to $\bar g_{\mu\nu}$, so that $\bar\na_\la  \bar
g_{\mu\nu}=0$ (but not for $\hat \ga_{\mu\nu}$).
Fixing  further the particular gauge $\zeta^\la=-\chi^\la$ one arrives at
 the quartet field disappearance:
\be
\chi^\mu\to\chi'^\mu=0,
\ee
in the favour of the  metric  field  redefinition:
\bea
 h_{\mu\nu} &\to&  h'_{\mu\nu}= h_{\mu\nu} - (\bar g_{\mu\la} \bar \na_\nu
\chi^\la +  \bar g_{\nu\la} \hat \na_\mu \chi^\la) ,\nn\\
 h &\to&h'  \equiv \bar g^{\mu\nu}  h'_{\mu\nu}=h-2\bar \na_\la \chi^\la.
\eea
It follows henceforth that after such transformations the whole Lagrangian $L_G$
(and, in particular, $L_g$) gets dependent  only on $h'_{\mu\nu}$. Due to   
$\ga_{\mu\nu}\to  \hat \ga_{\mu\nu}$,
the  fully dynamical theory in the  WF limit  reduces  thus  to the 
semi-dynamical model,  with the  redefined metric field $h'_{\mu\nu}$ 
and  the given   nondynamical  affine connection $\hat
\ga^\la_{\mu\nu}(\hat\ga_{\ka\la})$.  In particular, presenting $\si$~as 
\be 
\si= -\frac{1}{2}\ln |\det (\Si^{\mu\nu})\hat Q^2|
\ee
one  gets 
\be 
\si=\ln\sqrt{-\bar g}/\sqrt{-\hat {\ga}} +\frac{1}{2} h',
\ee
as in the minimal semi-dynamical model, with $\bar g\equiv  \det(\bar
g_{\mu\nu})$ and $\hat {\ga} \equiv 
\det(\hat {\ga}_{\mu\nu})=- \hat Q^2$. 
The  similar  consideration  remains  true  with account for a matter
Lagrangian, $L_m$.

This  presents  an implementation of the Higgs mechanism for the
extended gravity: the gauge $\zeta^\la=-\chi^\la$  totally 
absorbs $\chi^\la$  in the  favour of the  four
additional gravity d.o.f.'s contained in 
$h'_{\mu\nu}$,  chosen as   a new dynamical variable. 
Due to the gauge invariance, the WF theory in an arbitrary gauge, a WF remnant
of $L_F$,  still  describes 
the same  ten   d.o.f.'s (the  six tensor ones originating ultimately  from 
the metric field $h_{\mu\nu}$) and the four scalar and vector ones (originating
from the quartet  $\chi^\mu$)
incorporating, possibly, the ghosts. Further    clarifying  the
nature  of these   d.o.f.'s  (propagating or not, ghost or not) 
depends on a residual gauge invariance/relativity  determined,  in
turn,  by  the Lagrangian  parameters.  This question is addressed to  below.

\subsection{Linearized approximation }

Consider   now the  simplest   version of the WF theory, when the  metric and
quartet  backgrounds are globally  flat, presenting evidently a   solution
to the vacuum  FE's.  At least, this can be treated as  an approximation, in
neglect by the space-time curvature,  in a
region around a nonsingular space-time  point.
  Moreover, choose 
the distinguished  observer's  coordinates for the background,  $\hat {q}^\al
=\de^\al_a \hat {Q}^a$  ($\hat {Q}^a=\de_\al^a \hat {q}^\al$),  where there 
simultaneously fulfills
\be 
\hat {Q}^a_\al   = \partial_\al \hat {Q}^a=\de^a_\al, \ \ \bar g_{\al\beta}=\hat
{\ga}_{\al\beta} =
{\eta}_{\al\beta},
\ee
with ${\eta}_{\al\beta}$  the Minkowski symbol, by means of which the indices
are operated.
In view of   $ \Si^{\al\beta} = {\eta}^{\al\beta} -  h'^{\al\beta}$,  
where  
\bea 
h'_{\al\beta}&=&h_{\al\beta}-(\partial_\al\chi_\beta+ \partial_\beta
\chi_\al),\nn\\
h' &\equiv& \eta^{\al\beta} h'_{\al\beta}=h -2\partial_\ga \chi^\ga,
\eea
it follows that
\bea
\tilde\Si^{\al\beta} &=& -(h'^{\al\beta} -  \frac{1}{4}{\eta}^{\al\beta} h' ),
\nn\\
 \Si_0&=& -  h',  \  \  \si= \frac{1}{2} h'.
\eea
Note, that $h'_{\al\beta}$ remains invariant under the gauge transformations
(\ref{gt}), while $\chi^\al$ disappears at $\zeta^\al=-\chi^\al$.
Due to,   $\bar \Ga_{\al\beta}^\ga= \hat  \ga_{\al\beta}^\ga=0$, one also gets
\be
B^\ga_{\al\beta}= \frac{1}{2}{\eta}^{\ga\de}(\partial_\al h'_{\beta\de}+
\partial_\beta h'_{\al\de} - \partial_\de h'_{\al\beta}).
\ee
Under  the two  background  connections being
zero, the  (nonlinear) WF limit is nothing but LA.

Choose  a complete (up to the total derivatives) set
of the second-order partial kinetic operators in an obvious notation as follows:
 
\bea
 {K}_t=(\partial_\ga h'_{\al\beta})^2, 
&&
 {K}_s= (\partial_\al h')^2   ,    \nn\\
 {K}_v=  ( \partial^\beta h'_{\al\beta})^2 
,&&  {K}_{x}= \partial^\al h'_{\al \beta}\partial^\beta h' .
\eea
Expanding the Lagrangian  
up to the second order in the redefined metric field and eliminating ${K}_t$ in
the favour of $L_g$
one gets
\be\label{tvs} 
  L_G =  (1+\ve_t)L_g + \De K_{vs}  -\frac{1}{8}m_t^2\Big ((h'_{\al\beta})^2
-h'^2\Big) - \frac{1}{8}m_s^2 h'^2 ,
\ee
where
\be
m_s^2\equiv m_0^2 \mp m_x^2+m_\si^2
\ee
and 
\bea 
L_g&=&\frac{1}{8}( {K}_t  - 2{K}_v   + 2{K}_{x} -   {K}_s),\nn\\ 
\De K_{vs}  &=& \frac{1}{8}(   \ve_v {K}_v+ \ve_{x}  {K}_{x}+ 
\ve_s  {K}_s ) .
\eea
In what follows, we will intently preserve   the
primes  to stress that  the metric field at hand is  the redefined one 
absorbing   $\chi^\al$.

The five-to-four projection for the partial constants is as follows:
\bea
\ve_t&=& 3\ve_4-\ve_5 , \ \ 
\ve_v=4(\ve_2 +\ve_4)  ,  \nn\\
\ve_{x}&=&   -   2(\ve_2-\ve_3 +3\ve_4-\ve_5)  , \nn\\
\ve_s&=&      \ve_1+\ve_2-\ve_3 +3\ve_4-\ve_5.
\eea
Inverting,  one gets 
\bea
\ve_1&=&\ve_s+\frac{1}{2}\ve_x  ,\ \ 
\ve_2=\frac{1}{4}\ve_v -\la,  \nn\\
\ve_3&=&  \frac{1}{4}\ve_v+\frac{1}{2}\ve_x+\ve_t  -\la, \ \ 
\ve_4= \la , \nn\\
\ve_5&=&-\ve_t +3\la    ,
\eea
where $\la $ is a small arbitrary  parameter. The latter may equivalently be
redefined in the favour of a linear combination of itself and $\ve_v$,  $\ve_x$,
and $\ve_t$. The presence of such an undetermined  parameter results in the
partial decoupling of the full nonlinear theory and its WF limit, allowing one,
in principle,   to vary the former under choosing  the latter.
Clearly, the  pair ($\ve_1$, $\ve_s$) constitutes a  closed  uniquely
invertible subset of all the parameters, with the pure-scalar operator ${K}_1$
being in this sense special.  Its minimal modification without changing
LA  may be obtained by adding in the full nonlinear theory the  terms
proportional to $\la$.

Though  after choosing  the  gauge  $\zeta^\al=-\chi^\al$ the coordinates
in LA are already fixed leaving,  generally, the ten independent variables
$h'_{\mu\nu}$, $L_G$  may still  possess in LA by a residual gauge invariance
further reducing  the number of the d.o.f.'s.  Namely, under~Diff's
\be\label{Dh}
{\cal D}_{{\varphi}} h'_{\al\beta} =\partial_\al
{\varphi}_\beta+\partial_\beta {\varphi}_\al, \ \  {\cal D}_{{\varphi}} h' =
2\partial_\ga
{\varphi}^\ga
\ee
one gets (up to the total derivatives)
\bea\label{DO}
 {\cal D}_{{\varphi}}  {K}_t&=& -2( \partial_\al{\varphi}_\beta +
\partial_\beta{\varphi}_\al) 
\dal h'^{\al\beta}, \nn\\
{\cal D}_{{\varphi}}    {K} _v&=& -  (\partial_\al{\varphi}_\beta+  
\partial_\beta{\varphi}_\al)  \dal h'^{\al\beta} -
 \partial_\ga {\varphi}^\ga  \partial_\al\partial_\beta h'^{\al\beta},\nn\\
{\cal D}_{{\varphi}}     {K}_{x}  &=& -\partial_\ga {\varphi}^\ga( \dal h'
+\partial_\al\partial_\beta h'^{\al\beta}),\nn\\
{\cal D}_{{\varphi}}  {K}_s&=&-2\partial_\ga {\varphi}^\ga \dal h' ,
\eea
where $\partial^2\equiv {\eta}^{\al\beta}\partial_\al\partial_\beta$, as well
as 
\bea
{\cal D}_{\varphi} (h'_{\al\beta})^2&= &2 ( \partial_\al{\varphi}_\beta +
\partial_\beta{\varphi}_\al) h'^{\al\beta } ,\nn\\
{\cal D}_{\varphi} h'^2&=& 4 \partial_\ga{\varphi}^\ga h'.
\eea
It follows henceforth that   $L_g$ is by the very construction  
GDiff-invariant, ${\cal D}_{{\varphi}}L_g=0$, whereas the rest of $L_G$ 
(\ref{tvs})
is always GDiff-variant.  Most generally thus,  LA 
possesses by  no residual Diff's.  Inclusion of  the four
scalar components in the original Lagrangian 
results  in LA   in the four more  gravity d.o.f.'s.
With no  residual  Diff's,  all the ten d.o.f.'s   are thus  propagating. This
may cause some theoretical problems related with  the appearance of the ghost
vector graviton, e.g.,  the classical
instabilities~\cite{Alv}.
To abandon this, one may   impose {\em ab initio}  some  restrictions on the
parameters of $L_G$, discussed below.

\subsection{Residual transverse-diffeomorphism   invariance }

To consistently exclude the vector graviton let us  require 
\be\label{TD} 
\ve_v=4(\ve_2+\ve_4)=0, \   m_t=0.
\ee
While  the first requirement  is necessary,  the appearance of the
second  one is, in a sense,  sufficient (see, later on).\footnote{Moreover,
under $m_s\neq 0$, imposed  from the DS considerations,  the simultaneous 
fulfillment of $m_t\neq 0$ may result in ghosts~\cite{Alv}.}
The Lagrangian in LA  now becomes
\bea\label{s} 
  L_G& = & (1+\ve_t)L_g + \De K_{s}  -  \frac{1}{8}m_s^2 h'^2,   \nn\\
\De K_{s}  &=& 
\frac{1}{8}( \tilde\ve_{x}  {K}_{x}+ \tilde \ve_s  {K}_s ),
\eea
with   the same $\ve_t$, and the reduced partial constants as follows:
\bea 
\tilde\ve_{x}&=& 2(\ve_3-2\ve_4 +\ve_5), \nn\\
\tilde\ve_s&=&   \ve_1- ( \ve_3 -2\ve_4 +  \ve_5). 
\eea
It follows from (\ref{DO}) that the residual gauge symmetry of $L_G$ in LA  
increases under such the constraints  from no-Diff  up to the 
three-parameter TDiff: 
\be 
\mbox{\rm  TDiff} \  : \    \partial_\ga{\varphi}^\ga=0,  
\ee 
The two constraints (\ref{TD}) select the most general theory  possessing in LA
by no explicit problems.
On the one hand, the constraints  result in the appearance of TDiff. On the 
other hand,   TDiff ensures these 
constraints to be {\em natural}  in the  't Hooft's sense of increasing the
residual symmetry. Due to the  increased   symmetry, such the constraints
may survive  under the radiative corrections.\footnote{Except if 
TDiff is broken dynamically back to no-Diff, with 
$\ve_v\neq 0$ and  $m_t\neq 0$ reappearing due to the quantum effects.}

Under TDiff, to remove  in LA the arising gauge ambiguity 
one should  impose on $h'_{\al\beta}$   at the classical level   an extra  gauge
condition, 
e.g.,~\cite{Buch,Creutz}:
\be\label{gf} 
\partial_\al\partial^\ga h'_{\beta\ga}- \partial_\beta\partial^\ga
h'_{\al\ga}=0.
\ee
Decomposing $h'_{\al\beta}$ as
\be 
h'_{\al\beta}  = \tilde h'_{\al\beta}  + \frac{1}{4} {\eta}_{\al\beta} h',
\ee
with $ \tilde h'\equiv {\eta}^{\al\beta} \tilde  h'_{\al\beta}=0$, one sees 
that $h'$ is
unrestricted  by the gauge condition. Moreover, the  condition implies
$\partial^\ga \tilde h'_{\al\ga}=\partial_\al \tilde f$, with $\tilde f$ an
arbitrary scalar
function, indicating the fulfillment of  just three independent restrictions on
$ \tilde h'_{\al\beta}$. At the quantum level, one should add in LA  the
respective 
gauge fixing Lagrangian
\be\label{gf'} 
L_F=\al (\partial_\al\partial^\ga h'_{\beta\ga}-
\partial_\beta\partial^\ga
h'_{\al\ga})^2,
\ee
with $\al$ a dimensionless  gauge parameter.   At $\al \to \infty$, $L_F$
ensures the classical restriction (\ref{gf}).\footnote{ For a
higher-derivative  Lorentz-symmetric gauge, cf.~\cite{Alv}. Alternatively,   to
eliminate just three vector components, not touching the scalar one,  there
could to  used a Lorentz-non-symmetric gauge. Due to the residual 
gauge invariance this should not violate the Lorentz symmetry.}
The gauge fixing Lagrangian  additionally  eliminates out of $L_G$  in LA  three
d.o.f.'s leaving seven independent ones,  compared to ten  under no-Diff  and
six under GDiff (realized at $\tilde \ve_x=\tilde \ve_s=0$ ). The three more 
gravity components, the vector ones,   will appear  only beyond
LA. But having no quadratic propagator they will  just modify the higher-order 
vertices by means of the contact  interactions.\footnote{Such an  extra linear 
gauge is an artifact of LA with TDiff. In a general case, 
with all the ten d.o.f.'s in play, this gauge  should be abandoned.} 

While the  most general  {\em natural}  TDiff case with $\tilde \ve_x\neq 0$ 
deserves  a special consideration, let  for  simplicity 
$\tilde\ve_x=0$,  resulting in  $\tilde \ve_s=\ve_1$. This  implies
one more restriction, $\ve_3-2\ve_4+\ve_5=0$,  on the Lagrangian parameters.
However, not increasing the symmetry such a
restriction is not natural  in the 't Hooft's sense.
In the  minimal  case, one gets
\be 
L_G= (1+\ve_t)  L_g+\frac{1}{2}(\partial_\al \vsi)^2 -\frac{1}{2}M_s^2
\vsi^2,
\ee
where we have introduced  a  (dimensionfull)  physical systolon  field 
$\vsi\equiv \ka_s h'/2$, 
$\ka_s \equiv \ve_1^{1/2}  \ka_g $ ($\ka_g=1$), with  a  physical mass 
$M_s\equiv m_s \ka_g/\ka_s =m_s/\ve_1^{1/2}$.
Such a particular case   presents   a consistent quantum field
theory, unitary and free of ghosts~\cite{Buch}--\cite{Creutz1}.  It  describes
the massless two-component transverse-tensor graviton and its  massive scalar
counterpart. Neglecting by  $m_s$, one can
estimate from the anomalous asymptotic rotation   velocities in galaxies,
$v_\infty$,  that  $\ve_1^{1/2}\sim v_\infty/c\sim
10^{-3}$~\cite{Pir1a,Pir1}. It is seen  that 
the limit $\ve_1\to 0$ at a finite  mass parameter $m_s$ would  correspond
to a non-propagating heavy systolon $\vsi$, with its  physical mass 
$M_s \to\infty$. Thus the account for $m_s\neq 0$  could be of importance.

Altogether, the  minimal  version of QMG  given  by  the
two independent parameters $\ve_1>0$  and 
$m_s\neq 0$,   in principle,  suffices  to encompass both  the gravity  
DM and~DE.    The account for $\ve_3\neq 0$,  resulting, in
particular,  in the kinetic mixing $\tilde \ve_x=2\ve_3 $,   would extend the
range of the phenomenological possibilities. Though the tensor gravity  in LA 
remains  the same as for GR  (up to the overall normalization), 
the additional account for $\ve_4, \ve_5\neq 0$  in the different combinations
(under $\ve_2=-\ve_4$) would modify the tensor gravity  
 in the full nonlinear theory beyond GR, 
extending thus the  observational possibilities for DS even
further.
At last, assumption for $\ve_v\neq 0$ ($\ve_2\neq - \ve_4$) would ruin the
residual Diff invariance in  LA  up to no-Diff (being, as it was
mentioned, problematic). In this case, the number of the propagating
d.o.f .'s in LA  equals to ten, the vector  part of them
being ghosts.\footnote{On the contrary, if only $\ve_t\neq 0$, then the
residual gauge symmetry in  LA  {\em naturally} increases  up to 
GDiff, leaving just six propagating d.o.f.'s, similarly to  GR (though under a
modified  full nonlinear~theory).}

\section{Conclusion}

The given consideration shows that  the quartet-metric (QM)  GR (or, otherwise,
QMG)
may  well serve as the theory of  the unified  gravitational  DM and DE (the
gravity DS).  Under the natural restriction on  parameters,  the
theory   in LA,  being  unitary and free of ghosts, as well as
 the classical instabilities,   consistently comprises  the massless tensor
graviton and its massive scalar counterpart, the systolon,  as the DM particle.
The sufficient   abundance of  the free
parameters in the   full nonlinear theory and the  partial decoupling of the
latter  from  its WF  limit  noticeably   extend the prospects  for the
manifestations of the gravity DS  in the various phenomena  at the
drastically different scales. 
Further theoretical study of  the theory,  as well as  its
observational verification/limitation,  is  urgent.
Accounting as the  effective field theory  of  gravity  beyond  GR 
for the influence of the vacuum, QM  GR  (under confirmation)  could 
eventually pave the  proper way  towards  a
(more) fundamental  theory of gravity and space-time.

\end{document}